\documentclass[9pt,twocolumn,twoside]{gsajnl}
\usepackage{multirow}
\usepackage{epstopdf}
\usepackage{makecell}
\newcommand{\zcis}{z_{\mathrm{cis}}}
\newcommand{\ztrans}{z_{\mathrm{trans}}}
\newcommand{\ztwas}{z_{\mathrm{TWAS}}}
\articletype{inv} 
\usepackage{xcolor}
\usepackage{colortbl}
\definecolor{tabhead}{HTML}{2C3E50}   
\definecolor{tabours}{HTML}{FCEBD2}   
\definecolor{tabalt}{HTML}{F4F6F8}    

\runningtitle{Disease-regulatory flux maps in CAD} 
\runningauthor{Mohammad \textit{et al.}}

\title{Mapping disease-regulatory flux through eQTL-based causal gene networks: a complex-trait framework applied to coronary artery disease.}

\author[1]{Gutama Ibrahim Mohammad}
\author[2]{Johan L. M. Bj\"orkegren}
\author[1,$\ast$]{Tom Michoel}
\affil[1]{Computational Biology Unit, Department of Informatics, University of Bergen, Torm\o{}hlensgate 55, 5008 Bergen, Norway}
\affil[2]{Department of Medicine, Karolinska Institutet, Karolinska Universitetssjukhuset, Huddinge, Sweden}
\affil[$\ast$]{Corresponding author. \href{mailto:tom.michoel@uib.no}{tom.michoel@uib.no}}
\date{}

\begin{abstract}
\noindent A substantial proportion of inherited susceptibility to common complex diseases is mediated through tissue-specific regulatory variation, and the omnigenic model further hypothesizes that much of this risk arises from distant (trans) regulatory effects that are propagated via gene-regulatory networks from numerous regulatory genes onto a relatively small set of core genes. Two major challenges have hindered a direct empirical assessment of this model. First, transcriptome-wide association studies (TWAS) primarily assess gene expression that is regulated locally (in cis), and thus largely miss trans-acting effects. Second, trans regulation has so far been examined only in aggregate—either through its overall contribution to heritability or as an additional source of predictive power for expression—without identifying which specific regulatory genes transmit disease-associated signals to which target genes, or measuring the strength of each individual regulatory effect.

Here, we present a framework that decomposes each gene’s disease association into a local cis component and a set of trans components attributable to its upstream regulators, propagated along a directed causal gene-regulatory network whose edges are inferred using eQTLs as instrumental variables. For each gene, we estimate sparse Bayesian expression models with separate cis and trans terms, and we assess model performance using out-of-sample prediction. Propagating trans signals through the inferred network yields a disease-regulatory flux map: a directed, signed representation that quantifies the contribution of each regulator to the disease association of each target gene. Applying this framework to seven cardiometabolic tissues relevant to coronary artery disease, we demonstrate how genetic variation flows through the gene-regulatory network to ultimately affect disease risk. Outgoing regulatory influences from individual genes are directionally heterogeneous, whereas disease-associated genes tend to integrate convergent input from multiple upstream regulators; these convergent targets are enriched for cardiovascular-related biological processes. Consequently, each gene’s disease association can be reinterpreted as a detailed allocation of the disease signal among its contributing upstream regulators.
\end{abstract}


\dates{\rec{xx xx, xxxx} \acc{xx xx, xxxx}}

\begin{document}

\maketitle
\thispagestyle{firststyle}
\vspace{-13pt}

\section{Introduction}

Genome-wide association studies (GWAS) have discovered hundreds of associated loci for numerous complex traits, including coronary artery disease (CAD) \citep{aragamDiscoverySystematicCharacterization2022,howsonFifteenNewRisk2017,nelsonAssociationAnalysesBased2017,nikpayComprehensive1000Genomes2015}. These findings have facilitated downstream applications such as heritability estimation \citep{kheraGenomewidePolygenicScores2018}, genetic correlation analysis \citep{bulik-sullivanAtlasGeneticCorrelations2015} and polygenic risk prediction \citep{chenFastScalableEnsemble2024}

Despite these advances, GWAS have also highlighted substantial limitations. Linkage disequilibrium (LD) frequently obscures the specific causal variants that underlie observed association signals, and the majority of risk variants reside in noncoding regions. Consequently, the genes that mediate variant effects on disease are rarely identifiable from GWAS summary statistics alone \citep{wainbergOpportunitiesChallengesTranscriptomewide2019, uffelmannGenomewideAssociationStudies2021}. In addition, most associated variants are distributed widely across the genome, exhibit small individual effect sizes, and genome-wide significant loci account for only a modest fraction of the expected heritability, contributing to the so-called ``missing heritability'' problem \citep{manolioFindingMissingHeritability2009}.

The omnigenic model offers a conceptual framework for these observations by partitioning genes into a relatively small set of core genes, whose expression directly influences the trait, and a large peripheral set of regulatory genes that affect the trait indirectly by perturbing core gene expression in trans through highly interconnected regulatory networks \citep{boyleExpandedViewComplex2017,liuTransEffectsGene2019}. Because most genes are subject to trans regulation and regulatory networks are densely connected, the omnigenic model predicts that the majority of heritability is mediated through trans-regulatory pathways, and that peripheral genes with coordinated effects on many core genes can themselves exhibit detectable marginal genetic association signals \citep{liuTransEffectsGene2019}.

Transcriptome-wide association studies (TWAS) test whether genetically predicted expression of a gene associates with a trait \citep{gamazonGenebasedAssociationMethod2015, gusevIntegrativeApproachesLargescale2016, barbeiraIntegratingPredictedTranscriptome2019}. As conventionally implemented, the expression predictor is built from cis-eQTL variants only, a fundamental restriction relative to the omnigenic model: a cis-only TWAS can detect core genes with cis-regulated expression, but it is blind to the trans channel that the omnigenic model places at the centre of complex-trait architecture, and, because predicted expression is correlated among co-regulated genes, an association at one gene may instead reflect regulatory signal transmitted from others \citep{wainbergOpportunitiesChallengesTranscriptomewide2019}. Efforts to account for the trans contribution have taken two routes, neither of which resolves the regulatory wiring. Heritability-partitioning methods report aggregate, genome-wide quantities: how much trait heritability falls in regulatory elements (S-LDSC, across functional annotations) or how much of expression heritability is trans-acting (the omnigenic trans-$h^2$ estimator)
\citep{reprogenconsortiumPartitioningHeritabilityFunctional2015,liuTransEffectsGene2019}; they name neither the regulators that relay the signal nor the cis/trans split of an individual gene's association. Trans-aware association methods incorporate distal eQTLs or gene-set effects to recover power
\citep{bhattacharyaMOSTWASMultiOmicStrategies2021,liuGBATGenebasedAssociation2020,wangTranseQTLMappingGene2024}; some identify local mediators or candidate upstream genes, but none reconstructs a causal regulatory network or decomposes a target gene's association into its cis and trans parts. A separate line of TWAS fine-mapping methods takes the opposite stance, treating correlated predicted expression (induced by shared eQTLs and LD among co-regulated genes) as a confound to be removed so as to isolate a small set of clean cis-causal genes \citep{zhaoAdjustingGeneticConfounders2024,mancusoProbabilisticFinemappingTranscriptomewide2019}. Our aim is complementary to this fine-mapping: rather than removing the propagated trans cascade to expose cis-causal genes, we make that cascade itself the object of
study. Individually, trans-eQTLs are weak and difficult to detect \citep{liuTransEffectsGene2019,albertGeneticsTransregulatoryVariation2018,vosaLargescaleCisTranseQTL2021}; we reason, however, that their aggregate effect on a target gene is largely funnelled through its upstream regulators, which are themselves under detectable cis-genetic control. A predictor that propagates a regulator's cis-eQTLs to its downstream targets through an inferred regulatory network can therefore both recover a usable trans signal and attribute it to those upstream regulators.

The omnigenic model anticipated this object: high-quality trans-regulatory networks and trans-eQTL information, combined with GWAS effect sizes, would describe how genetic effects flow through the regulatory network---which the model noted was precisely what was missing \citep{liuTransEffectsGene2019}. Here we construct it. We introduce the \emph{disease-regulatory flux map}, built in three steps within each tissue: (i)~we reconstruct a causal, cis-anchored gene-regulatory network over the full set of expressed genes, orienting every edge by the source gene's own cis-eQTL; (ii)~we fit a channel-aware Bayesian expression model for every gene, predicting it from its own cis-eQTLs (the cis channel) and from the cis-eQTLs of its upstream regulators propagated along the network edges (the trans channel) \citep{mohammad_predicting_2024}; and (iii)~we carry both channels through a summary-statistic TWAS whose association statistic separates additively into a cis component and a trans component.

This framework offers three primary contributions. First, it separates each gene’s disease association into two parts: one arising from the gene’s own local (cis) variation, and another propagated through the network from its upstream (trans) regulators. To our knowledge, no prior work has carried out this specific type of decomposition, which is essential for the two subsequent advances. Second, it assigns the network-derived component back to the individual regulators that convey it: the trans signal is expressed as an exact sum of signed, per-regulator contributions, so that every edge from a regulator to a disease-associated gene is given a signed weight. This ensures that genes whose association is purely trans-driven with no detectable cis effect remain linked to disease via the regulatory network. Third, by aggregating over all disease-associated genes and their regulators, these signed edges produce a directed \emph{disease-regulatory flux map} in each tissue: a weighted representation of how much disease association each regulator transmits to each target gene, thereby converting the omnigenic model’s structural assertion—that disease signal travels through regulatory networks onto core genes—into an empirically quantifiable entity. This map, in turn, allows us to rank and prioritize mediators not only by statistical significance, but also by whether they are central hub genes at the core of regulatory flux, integrating and redistributing effects; members of a broad periphery of regulators whose small individual influences sum to a large overall impact; or convergence nodes where many modest inputs combine to produce substantial variation in trait risk. We anticipate that this systems-level prioritization approach will substantially aid downstream efforts such as therapeutic target identification. In addition, we provide a data-agnostic, configuration-driven pipeline that can be used to construct a regulatory flux map for any complex disease, given an appropriate reference cohort and GWAS summary statistics.

This map also ranks genes by how much disease signal they broadcast to others, thereby revealing regulators that extensively relay disease signal without themselves achieving genome-wide significance.

We construct and use this framework for coronary artery disease (CAD) in seven cardiometabolic tissues from STARNET \citep{franzenCardiometabolicRiskLoci2016}, generating one disease-regulatory flux map per tissue and analyzing each map along three axes: (1) beyond its own direct association, how much disease signal does it absorb from its regulators, and how much does it transmit onward to disease genes? (2) coordination—when a gene sends signal to many disease genes, do these outgoing contributions push in a consistent direction, or do they oppose and neutralize each other? (3) convergence—when a disease gene is influenced by many upstream genes, do these incoming contributions share the same sign, such that the disease gene is coherently targeted? Axes (2) and (3) use the same coherence metric applied in opposite directions, and together they assess whether CAD’s trans architecture exhibits the structural patterns predicted by the omnigenic model—namely, the channelling and convergence of signal onto disease genes.

\section{Materials and Methods}

\begin{figure*}[htbp]
\centering
\includegraphics[width=\textwidth]{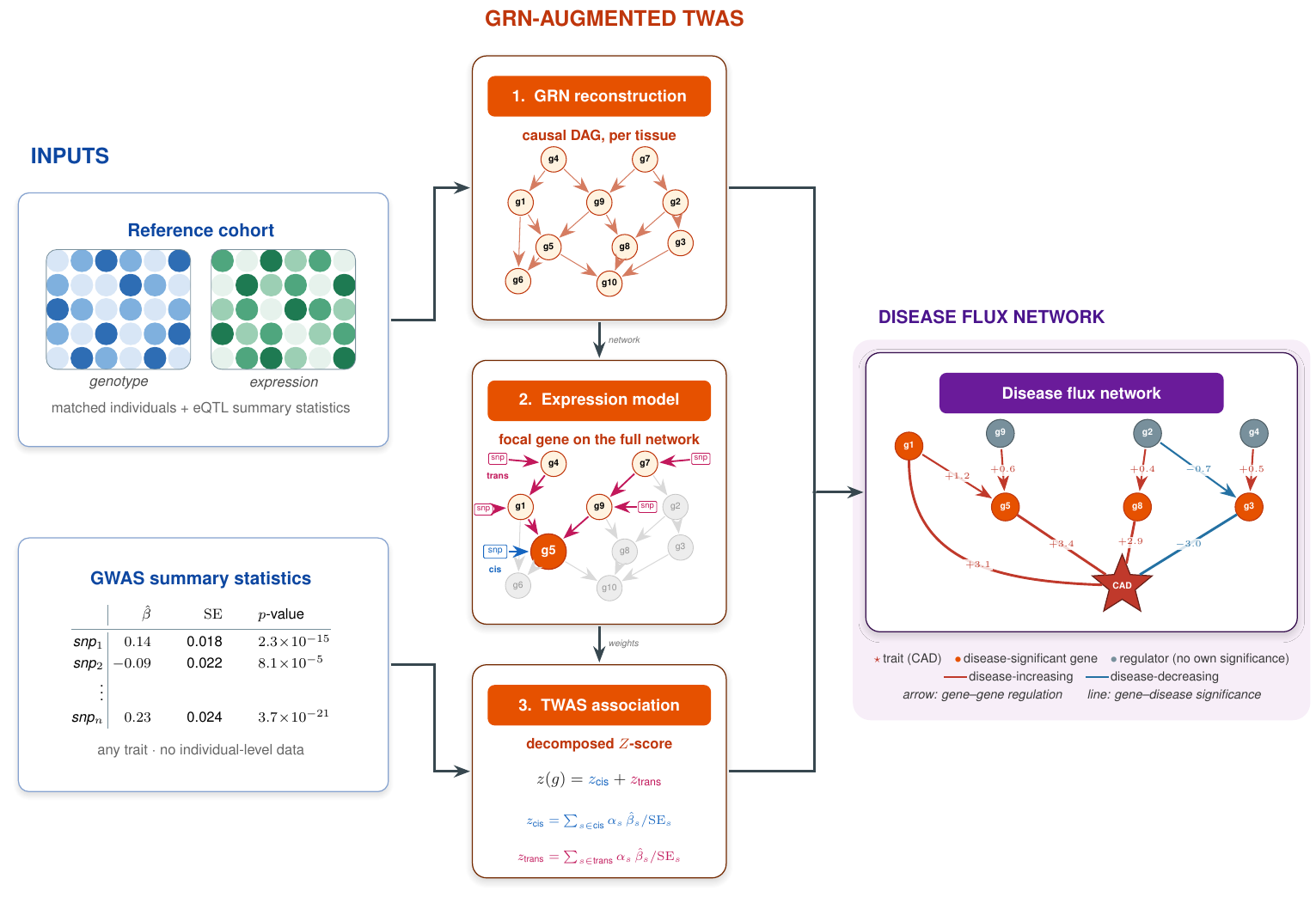}
\caption{Overview of the disease-regulatory flux framework. \textbf{Inputs:} a reference cohort of matched genotype and expression (and the cis-eQTL summary statistics derived from the same reference) and GWAS summary statistics for any trait. \textbf{GRN-augmented TWAS:}
(1)~a causal gene regulatory network is reconstructed per tissue as a directed acyclic
graph over the full set of expressed genes; (2)~a channel-aware model predicts each  gene
(example highlighted, g5) from two channels defined on this network, its own cis-eQTLs (cis) and the cis-eQTLs of its upstream ancestors propagated along the edges (trans); (3)~combined with the GWAS, this yields a summary-statistic TWAS whose $Z$-score splits additively into a cis and a trans component, $z(g)=\zcis+\ztrans$. \textbf{Disease flux network:} the output is a subset of the full GRN that mediates genetic effects to disease. Directed gene$\to$gene edges carry the signed flux a regulator delivers to a gene (red, toward disease risk; blue, away; the value is labelled on the edge); a gene that is itself disease-significant is joined to the trait node (star) by an undirected significance edge. Genes that only relay trans signal to disease genes, with no significance of their own (grey), have no edge to the trait.}
\label{fig:pipeline}
\end{figure*}
The framework comprises three major components (Fig.~\ref{fig:pipeline}). The \emph{input} component requires a reference cohort with matched individual-level genotypes and tissue-specific gene expression, the cis-eQTL summary statistics derived from that same reference, and independent GWAS summary statistics for a complex trait of interest. The \emph{GRN-augmented TWAS} component reconstructs a directed, cis-anchored gene regulatory network for each tissue over the full set of expressed genes, orienting every edge by the source gene's own cis-eQTL; fits a channel-aware expression model for every gene that integrates the gene's own cis-eQTLs with the cis-eQTLs of its upstream regulators propagated along the network edges; and computes a TWAS association statistic that decomposes additively into a cis and a trans component. The \emph{disease-regulatory flux map} component then attributes each gene's trans component to the specific upstream regulators that deliver it: the trans component is an exact sum of signed per-regulator contributions, so every regulator-to-gene link carries a signed weight, and these links assemble into one signed flux map per tissue. Routine preprocessing steps are carried out by a series of intermediate stages. These include genotype and expression harmonisation, per-gene cis-SNP selection, LD pruning, and construction of trans features by walks on the GRN.

\subsection{Data}
For our primary analyses, we used the STARNET cohort \citep{franzenCardiometabolicRiskLoci2016}, which provides matched genotyping and RNA-seq data across multiple cardiometabolic tissues. We focused on seven tissues: atherosclerotic aortic root (AOR), blood, liver (LIV), mammary artery (MAM), subcutaneous fat (SF), skeletal muscle (SKLM), and visceral fat (VAF). Discovery GWAS summary statistics for CAD were taken from the CARDIoGRAMplusC4D meta-analysis \citep{aragamDiscoverySystematicCharacterization2022} and harmonised to the genotype alternate (ALT) allele. Gene expression values were drawn from the processed STARNET expression matrices of \citet{franzenCardiometabolicRiskLoci2016}, which had already undergone GC-content and sequencing-depth normalisation (EDASeq), adjustment for known technical and biological covariates, outlier removal, and rank-based quantile normalisation. We used these processed matrices as provided, without applying any additional residualisation. 

\subsection{GRN reconstruction}
To reconstruct tissue-specific gene regulatory networks (GRNs) from reference genotype and transcriptome data, we used the pairwise causal inference framework \textit{\href{https://github.com/lingfeiwang/findr}{Findr}} \citep{wangEfficientAccurateCausal2017, wangHighDimensionalBayesianNetwork2019}. In this framework, GRNs are inferred by first identifying, for each gene $A$ with a significant cis-eQTL $E$ (defined as the variant within 1 Mb of the transcription start site of $A$ that has the smallest association $p$-value, among the tissue's cis-eQTLs called significant at FDR~$\le 0.05$), an instrumental variable for that gene. Any gene possessing such a cis-eQTL instrument can serve as a regulator; we refer to these regulatory genes as \emph{A-genes}, and to all other expressed genes—which are potential regulatory targets—as \emph{B-genes}. Using $E$ as an instrumental variable for the A-gene, the framework then evaluates the association between $A$ and every candidate B-gene. A set of six likelihood-ratio tests based on the instrument $E$ jointly discriminates direct regulation $A\!\to\!B$ from reverse causation, confounding, and pleiotropy. The FDR-adjusted posterior probabilities from these tests are aggregated into a single edge score $P(A\to B)$. Edges that pass this scoring step are assembled into a maximum-weight directed acyclic graph, whose acyclic structure guarantees that the recursive trans-prediction from upstream network regulators (described below) terminates. Please refer to original papers \citep{wangEfficientAccurateCausal2017, wangHighDimensionalBayesianNetwork2019} for details of gene regulatory network reconstraction.

\subsection{Expression models}

With the regulatory network established, we model each gene’s expression as arising from two genetic channels: the gene’s own cis-eQTLs and the cis-eQTLs of its upstream regulators. The gene’s \emph{cis} channel is defined as its significant cis-eQTLs after LD pruning. Pruning proceeds by first selecting the lead cis-eQTLs and then  iteratively removing any SNP whose squared correlation ($r^2$) with an already retained SNP exceeds $0.7$. The \emph{trans} channel is formed from the LD-pruned cis-eQTLs of the gene’s GRN regulators, i.e. the cis-eQTLs of its ancestor genes up to two steps upstream in the DAG. From these, we retain at most the $200$ SNPs that exhibit the strongest correlation with the target gene in the training set. The two channels are disjoint at the SNP level: a variant that is both a regulator’s cis-eQTL and one of the target gene’s own LD-pruned cis-eQTLs is assigned to the cis channel only, so no SNP contributes to both channels. We apply no additional distance-based exclusion of trans SNPs that fall within the target’s cis window, because every regulator$\to$target edge is oriented by a causal test instrumented on the regulator’s own cis-eQTL, which discounts a direct variant-to-target effect; a regulator’s instrument therefore enters the trans channel as mediated signal rather than as disguised local cis signal.

For each gene, given mean-centered expression values $y$ across individuals and a genotype matrix $X$ (including cis SNPs, trans SNPs, or both), we fit a Bayesian linear model $y = X\beta + \varepsilon$ with a channel-aware shrinkage prior on the SNP effects $\beta$. We use a Bayesian approach for two main reasons. First, the full posterior distribution provides an estimate of out-of-sample predictive accuracy via leave-one-out cross-validation derived from a single model fit (see below), avoiding the need to refit the model once per held-out individual and never explicitly omitting any sample from training. Second, it naturally accommodates channel-aware shrinkage priors that heavily regularise the numerous weak instruments  while applying only mild penalties to true effects. We introduce separate global scale parameters for each channel so that the generally weaker trans channel can be pruned largely independently of cis. The pipeline implements three priors that trade off computational burden against posterior sparsity: a Bayesian ridge (Gaussian prior with one scale per channel), used as the fast default, and two heavier-tailed, more computationally demanding alternatives—a regularised (Finnish) horseshoe \citep{carvalhoHorseshoeEstimatorSparse2010, piironenSparsityInformationRegularization2017} and a BSLMM \citep{zhouPolygenicModelingBayesian2013} sparse-plus-polygenic prior. These heavier-tailed priors shrink the many non-informative instruments towards (or very close to) zero while leaving genuine regulatory effects only lightly shrunk. On this dataset, all three priors yield very similar predictive performance (Supplementary Section~S3). We present results based on the horseshoe prior, mainly because it concentrates each gene’s trans signal onto as few regulators as possible, producing the sparsest and most interpretable weight patterns and minimising shrinkage of true regulatory effects. This network-propagated trans predictor follows \citet{mohammad_predicting_2024}, who originally proposed it for predicting the genetic component of expression and demonstrated that its accuracy reflects underlying regulatory architecture: causal networks outperform correlation-based networks, whereas node-randomised networks perform worst.
Posterior inference was carried out with the No-U-Turn sampler \citep{hoffmanNoUTurnSamplerAdaptively2011} implemented in NumPyro \citep{binghamPyroDeepUniversal2018}, using $400$ warm-up iterations followed by $400$ sampling iterations. Posterior-mean effect sizes on allele dosage were then used in the downstream association analyses. For each gene, we fit up to three model variants—cis only, trans only, and cis$+$trans—and selected among them based on their expected log pointwise predictive density, estimated via Pareto-smoothed importance-sampling leave-one-out cross-validation (PSIS-LOO) \citep{vehtariPracticalBayesianModel2017}. We measured out-of-sample predictive accuracy using the leave-one-out coefficient of determination (LOO-$R^2$), and defined a gene as predictable if its selected model configuration achieved LOO-$R^2 \ge 0.01$. The prior specifications, including the hyperpriors and channel-specific scales, the non-centred parameterisation, and the PSIS-LOO estimator are described in Supplementary Section~S2.

For each tissue, both the regulatory network and the expression prediction models are inferred from the same reference cohort: the network is estimated once per tissue using all available samples, and each gene’s predictive model is fit on this same set of individuals, with out-of-sample accuracy assessed via PSIS-LOO. Relying on the same cohort for both network reconstruction and predictor training raises the possibility that feature selection on the full dataset could induce data leakage that, in turn, affects downstream associations. However, the trait is not used at either of these stages: it enters the analysis only later, via external GWAS summary statistics in the association step (Eq.~\eqref{eq:twasz}). As a result, all trait associations are, by construction, out of sample, analogous to standard cis-TWAS, where eQTL-based weights are combined with an independent GWAS. Any potential leakage is therefore confined to the prediction stage and does not extend to the association step; we explicitly examine and delimit where such leakage can and cannot arise (Supplementary Section~S4).


\subsection{Summary-statistic TWAS and cis/trans decomposition}
For each gene $g$, given a chosen predictor with posterior-mean weights $w_{g,s}$ defined on the raw allele-dosage scale, we computed a PrediXcan-style summary-statistic TWAS statistic using trait GWAS results \citep{gusevIntegrativeApproachesLargescale2016},
\begin{equation}
\ztwas(g)=\frac{1}{\sigma_g}\sum_{s} w_{g,s}\,\mathrm{sd}_s\,z^{\mathrm{GWAS}}_s,
\qquad \sigma_g=\mathrm{sd}\!\big(X_{\mathrm{c}}\,w_g\big),
\label{eq:twasz}
\end{equation}
where $\mathrm{sd}_s$ is the standard deviation of SNP $s$’s allele dosage in the training sample (capturing the linkage-disequilibrium/variance component), $z^{\mathrm{GWAS}}_s$ is the GWAS $z$-score harmonised to the ALT allele, and $\sigma_g$ is the standard deviation of the genetically predicted expression computed over the centred genotype matrix $X_{\mathrm c}$, which serves to standardise the test statistic. Because the predictor incorporates two channels, the numerator of Eq.~\eqref{eq:twasz} can be exactly decomposed into cis and trans components: restricting the sum to cis-eQTLs of $g$ yields $\zcis(g)$, while restricting it to the trans SNPs yields $\ztrans(g)$, such that $\ztwas(g)=\zcis(g)+\ztrans(g)$. Gene-level significance was evaluated using a two-sided normal test on $\ztwas$, followed by Benjamini--Hochberg adjustment controlling the FDR at $<0.05$ \citep{benjaminiControllingFalseDiscovery1995}.  Because $\ztwas$ divides by the predicted-expression scale $\sigma_g$, a near-degenerate predictor with vanishing $\sigma_g$ could inflate $|\ztwas|$; we therefore applied two floors on $\sigma_g$ (in units of predicted expression). Genes with $\sigma_g<0.05$ were dropped as degenerate, and a gene whose selected predictor used the trans channel was not called significant when its $\sigma_g\le0.2$, since the trans signal of a small-$\sigma_g$ predictor is not trusted (the raw $p$-value is left unchanged).

\subsection{Disease-regulatory flux map}
The flux map converts the trans component of each disease gene into a directed, weighted representation of the upstream regulators that contribute to it. Per tissue, its nodes fall into two classes. The first is the \emph{disease genes}: genes that reach significance in the association stage (FDR~$<0.05$), whether through a local cis signal, through trans regulatory input, or both. The second is the \emph{upstream regulators} that deliver trans signal into those disease genes---each disease gene's hop-1 GRN parents and hop-2 grandparents---which need not themselves be transcriptome-wide significant. By construction of the trans channel, every trans SNP $s$ in Eq.~\eqref{eq:twasz} is a cis-eQTL for one or more of $g$'s GRN ancestors, so $\ztrans(g)$ breaks down exactly into contributions from those regulators. For a regulator $A$ that influences $g$ through the network, we define the
\emph{flux} it delivers as
\begin{equation}
\mathrm{flux}(A\to g)=\frac{1}{\sigma_g}\sum_{s\in\mathcal E(A)\cap\mathcal T(g)}
w_{g,s}\mathrm{sd}_s\,z^{\mathrm{GWAS}}_s
\label{eq:flux}
\end{equation}
where $\mathcal E(A)$ is the set of cis-eQTLs for $A$ and $\mathcal T(g)$ is the set of trans features for $g$. If a SNP instruments multiple ancestors, its contribution is split evenly among them (divided by the number of such ancestors), ensuring that the sum of all per-regulator fluxes recovers $\ztrans(g)$ exactly ($\sum_{A}\mathrm{flux}(A\to g)=\ztrans(g)$). Thus, this is a purely algebraic decomposition, not a fitted model. The quantity $\mathrm{flux}(A\to g)$ is signed: a positive value indicates that $A$’s genetic effects drive $g$’s expression in the risk-increasing direction for the trait, whereas a negative value indicates the reverse.

\paragraph{Coherence of flux.} The two directional readouts of the map (whether a disease gene's incoming fluxes agree in sign, and whether a regulator's outgoing fluxes push its targets consistently) use a single coherence metric applied in opposite directions. For any set of signed fluxes $\{f_i\}$ we define the coherence ratio
\begin{equation}
C=\frac{\bigl\lvert\sum_i f_i\bigr\rvert}{\sum_i\lvert f_i\rvert}\in[0,1],
\label{eq:coherence}
\end{equation}
which equals $1$ when the fluxes all share a sign (perfectly coherent) and approaches $0$ when they cancel. Applied to a disease gene $g$'s incoming fluxes $\{\mathrm{flux}(A\to g)\}_A$, it measures \emph{convergence}; since these sum to $\ztrans(g)$, its numerator is exactly $\lvert\ztrans(g)\rvert$. Applied to a regulator $A$'s outgoing fluxes $\{\mathrm{flux}(A\to g)\}_g$, it measures broadcasting \emph{coordination}. We evaluate each only at nodes with at least two edges (a disease gene with $\ge2$ regulators, or a regulator with $\ge2$ disease-gene targets), and call a disease gene \emph{convergent} when its incoming coherence exceeds $0.7$.

\subsection{Validation and replication}
We evaluated our results using three complementary lines of evidence: comparison of the inferred regulatory edges against curated interactions, replication of the association statistics in an independent GWAS cohort, and enrichment of prioritised genes for cardiovascular biology.

We compared the reconstructed edges against OmniPath \citep{tureiOmniPathGuidelinesGateway2016}, a database of curated directed molecular interactions, testing whether the overlap exceeds a degree-preserving null.

We assessed replication in FinnGen \citep{kurkiFinnGenProvidesGenetic2023} by re-running the same gene-based association procedure, changing only the GWAS summary statistics. As a replication phenotype we used the FinnGen coronary heart disease summary statistics, which is independent of the \citet{aragamDiscoverySystematicCharacterization2022} discovery GWAS, and we used FinnGen height summary statistics as a negative control. Summary statistics were drawn from FinnGen R11 (\url{https://www.finngen.fi/en/access_results}; endpoints \texttt{finngen\_R11\_I9\_CHD} and \texttt{finngen\_R11\_HEIGHT\_IRN}). We summarized replication in three metrics,\emph{Effect-size agreement}: the Pearson correlation between FinnGen and discovery $\ztwas$. \emph{Sign concordance}: the fraction of genes with the same sign in FinnGen and discovery, evaluated against a $50\%$ null using a one-sided binomial test. \emph{Independent significance}: the fraction of discovery-significant genes that also pass FinnGen’s Benjamini--Hochberg FDR threshold. Height serves as a negative control for all three: because genetically predicted expression is heritable and height is highly polygenic, an unrelated trait can still produce many TWAS associations. We therefore define replication in terms of whether the CAD map’s directions and effect sizes are reproduced, rather than by the raw number of significant genes, since for an unrelated phenotype these metrics should revert to chance.

As a further check beyond the height control, we tested whether the association signal genuinely follows the disease GWAS, since in principle a large, densely connected network combined with a highly polygenic GWAS might generate associations from structure alone. The only component derived from the disease GWAS is the SNP-level $z$; the network, expression weights, SNP standard deviations, and genotypes are all from STARNET and remain fixed. Because $\ztwas$ is linear in $z$, we disrupted only the SNP$\to$disease relationship by drawing an independent random sign for each SNP, $z^{\mathrm{GWAS}}_s\!\to\!\epsilon_s z^{\mathrm{GWAS}}_s$ ($\epsilon_s\in\{-1,+1\}$, i.i.d.\ across SNPs), and applying that one sign vector across all genes and tissues, which keeps the SNP set, LD structure, and $|z^{\mathrm{GWAS}}|$ intact; we then re-computed the entire map. Across 1000 such realizations, we re-applied the per-tissue Benjamini--Hochberg procedure at FDR~$<0.05$ and counted  significant gene–tissue pairs, comparing each observed count with its permutation null via a two-sided empirical $p$-value. Supplementary Section~S5 details the linear decomposition and explains why sign-flipping, rather than permuting $z$ across SNPs, defines the appropriate null.

Finally, we tested whether prioritised genes are enriched for cardiovascular biology against a curated cardiovascular gene universe assembled from Enrichr libraries \citep{chenEnrichrInteractiveCollaborative2013,kuleshovEnrichrComprehensiveGene2016} and, as a genetics-independent readout, mouse-knockout phenotypes, by one-sided Fisher's exact tests.

\section{Results}
\subsection{Gene Regulatory Network Summarization}

\begin{table*}[htbp]
\centering
\caption{Summary and reconstruction cost of the gene regulatory networks across seven STARNET tissues. Regulators are A-Genes (genes with a significant cis-eQTL); targets are any expressed gene receiving a directed edge; nodes are the unique genes appearing as a regulator or target. Out-degree is the number
of targets per regulator; in-degree, the number of regulators per target. $N$ is the cohort sample size and Genes the number of candidate target genes (all expressed genes). Wall-clock time and peak memory are for reconstruction (python Findr / libfindr, $8$ cores); because the tissues are
reconstructed independently they run concurrently as a job array, so the full set finishes in the same few minutes. }
\label{tab:grn}
\small\setlength{\tabcolsep}{4.5pt}
\begin{tabular}{lrrrrrrrrrrrr}
\toprule
\multicolumn{7}{c}{} & \multicolumn{2}{c}{Out-degree} & \multicolumn{2}{c}{In-degree} & \multicolumn{2}{c}{}\\
\cmidrule(lr){8-9}\cmidrule(lr){10-11}
Tissue & $N$ & Genes & Nodes & Edges & Regulators & Targets & mean & max & mean & max & Wall-clock & Peak mem.\\
\midrule
AOR   & 513 & 16{,}214 & 6{,}115  & 7{,}016  & 1{,}877 & 4{,}995  & 3.7  & 619     & 1.40 & 9  & 1.8\,min & 9.0\,GiB\\
Blood & 447 & 12{,}843   & 6{,}968  & 10{,}311 & 1{,}460 & 6{,}370  & 7.1  & 987     & 1.62 & 9  & 1.1\,min      & 5.5\,GiB\\
LIV   & 522 & 13{,}875 & 8{,}402  & 12{,}390 & 2{,}377 & 7{,}458  & 5.2  & 768     & 1.66 & 11 & 1.7\,min & 7.8\,GiB\\
MAM   & 528 & 15{,}458 & 7{,}801  & 10{,}656 & 2{,}084 & 6{,}810  & 5.1  & 621     & 1.56 & 10 & 2.7\,min & 8.8\,GiB\\
SF    & 548 & 14{,}119 & 10{,}768 & 23{,}249 & 2{,}223 & 10{,}229 & 10.5 & 1{,}085 & 2.27 & 12 & 1.6\,min & 7.7\,GiB\\
SKLM  & 511 & 12{,}544 & 7{,}487  & 10{,}987 & 1{,}583 & 6{,}854  & 6.9  & 586     & 1.60 & 8  & 1.2\,min & 5.7\,GiB\\
VAF   & 508 & 14{,}964 & 12{,}093 & 29{,}922 & 2{,}205 & 11{,}624 & 13.6 & 1{,}226 & 2.57 & 16 & 1.8\,min & 8.0\,GiB\\
\bottomrule
\end{tabular}
\end{table*}

We summarize the reconstructed gene regulatory network for each tissue in Table~\ref{tab:grn}, including the number of edges, the counts of regulator and target nodes, and the minimum and maximum in- and out-degrees; the full edge lists are given in Supplementary Table~S1. We retained edges at a global FDR of $0.15$, a deliberately permissive threshold chosen to keep the network dense enough for trans signal to reach many targets while individual edges stay confident: even the weakest retained edge carries a posterior probability $P\gtrsim0.78$ of being a true regulatory interaction. The two adipose depots (subcutaneous, SF; visceral, VAF) produce substantially denser networks than the other five tissues (mean out-degrees $10.5$ and $13.6$ versus $3.7$--$7.1$). This network density covaries closely with the background co-expression of the expression matrices (Pearson $r=0.90$; Supplementary Section~S1, Fig.~S1), which is highest in the adipose tissues.

This pairwise, instrument-anchored test scores each source--target pair independently from the ource gene's cis-eQTL, so reconstruction is a set of per-pair likelihood-ratio tests that parallelise trivially and scale roughly linearly in the number of pairs. Alternative approach such as Constraint-based structure
learners for example (the PC algorithm \citep{spirtesAlgorithmFastRecovery1991,buhlmannVariableSelectionHighdimensional2010}), by contrast,
search the joint conditional-independence structure with conditioning sets that grow combinatorially---prohibitive at transcriptome scale ($\sim$$10^4$ genes)---and do not exploit the eQTL as a causal anchor to orient edges. On the seven STARNET tissues we analysed, despite evaluating $70$--$112$ million candidate
source--target edges per tissue, each network completed in $1$--$3$\,minutes of wall-clock time on $8$ cores (peak memory $\le 9$\,GiB per tissue; Table~\ref{tab:grn}). Because the tissues are reconstructed independently, all seven run concurrently as a job array and the full set finishes in the same few minutes of wall-clock time rather than seven times that.

\subsection{Channel-Aware Expression Model Summarization}
\begin{figure*}[htbp]
\centering
\includegraphics[width=\textwidth]{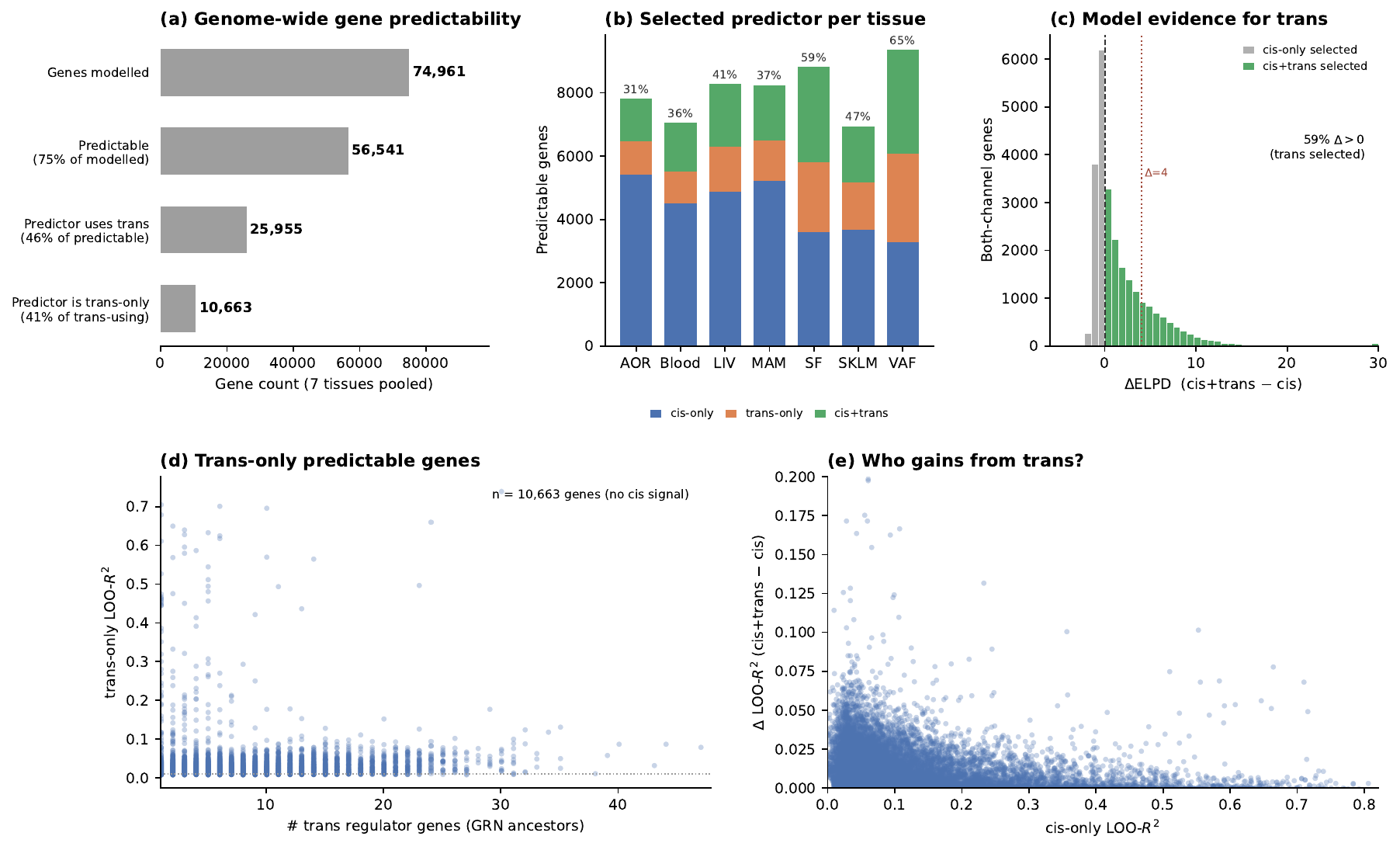}
\caption{Bayesian expression models across seven STARNET tissues. Every expressed gene is modelled as a prediction target from its own cis SNPs and from the cis-eQTLs of its GRN ancestor regulators up to two hops. For each gene the predictor uses the cis channel, the trans channel, or both, with the configuration chosen by PSIS-LOO
ELPD. A gene is predictable when its selected model reaches a cross-validated LOO-$R^2$ of at least 0.01. (a) Selection from all modelled genes to predictable genes pooled over the seven tissues, including the 10{,}663 trans only genes that have no cis signal of their own and become predictable through the network. (b) Composition of the selected predictor among predictable genes in each tissue, with the percentage that uses the trans channel shown above each bar. (c) Distribution across two channel genes of the leave-one-out evidence difference between the combined model and the cis only model. The combined model is selected where this difference is positive and is clearly favoured where it exceeds four. (d) Cross-validated LOO-$R^2$ of trans only predictable genes against the number of GRN ancestor regulators. (e) Gain in LOO-$R^2$ from adding the trans channel against the cis only LOO-$R^2$ for two channel genes.}

\label{fig:pred}
\end{figure*}

Across the seven STARNET cardiometabolic tissues, we fit a Bayesian horseshoe-prior model for \emph{each} gene in each tissues($74{,}961$ gene–models in total), using both the gene’s own cis-eQTLs (cis channel) and/or the
cis-eQTLs of its GRN regulators (trans channel). When both channels were present, we compared three model specifications—cis-eQTLs only, trans-eQTLs only, or a combined cis$+$trans eQTL model per gene—using leave-one-out predictive performance (PSIS-LOO $\mathrm{ELPD}$) and selected the model with higher ELPD. We then designated a model as suitable for downstream analysis (“predictable”) if its out-of-sample performance exceeded a field-standard threshold of $R^2 > 0.01$ (LOO-$R^2 \ge 0.01$).  Using this criterion, we identified $56{,}541$ genes ($75.4\%$) as predictable (Fig.~\ref{fig:pred}a; Supplementary Table~S2). Among these, in each tissue between 31–65\% had a trans channel, including both genes predicted solely from trans eQTLs (orange bar in Fig.~\ref{fig:pred}b) and genes predicted from the combined cis and trans eQTLs (green bar in Fig.~\ref{fig:pred}b).

To choose among cis only, trans only, and combined models for two-channel genes, we compared the expected log pointwise density (ELPD), which quantifies how likely unseen observations are under each specification using leave-one-out evaluation. Figure~\ref{fig:pred}c shows the distribution of ELPD differences between the combined cis$+$trans and the cis-only models. The combined model was preferred over cis only for $59\%$ of two-channel genes (positive ELPD), and strongly preferred ($\Delta\mathrm{elpd}>4$) for $20\%$.

For trans-only models, the trans signal is typically modest: most of their predictability falls below $R^2 < 0.1$, with an average of $R^2 \approx 0.03$ (Fig.~\ref{fig:pred}d). This contrasts with cis-only models, which have a higher average $R^2 \approx 0.11$. Importantly, the size of the parent set has little influence on the predictive performance of the trans channel; its contribution remains relatively consistent across different parent set sizes. In addition, for genes that have both channels and gain from including trans (Fig.~\ref{fig:pred}e), the largest improvements are seen in genes with poor cis predictability. This pattern suggests that some genes may lack strong cis regulation or have minimal cis effects, but are subject to multiple trans regulators whose combined influence is hard to detect given the available sample size.

\subsection{Network-Augmented TWAS Association}

\begin{figure*}[htbp]
\centering
\includegraphics[width=\textwidth]{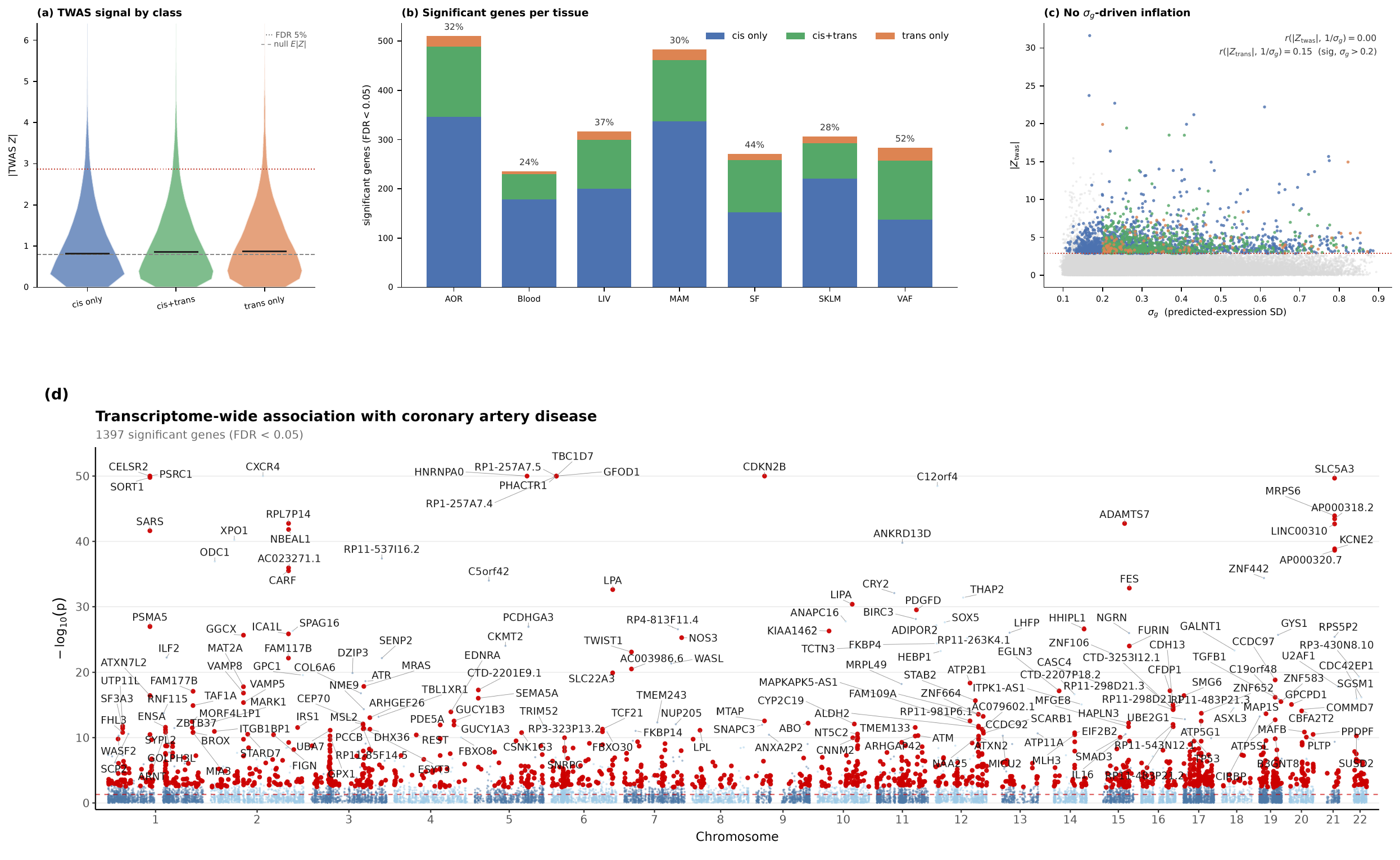}
\caption{CAD association across seven STARNET tissues. (a) Distribution of $|z_{\mathrm{TWAS}}|$ over all tested genes, grouped by the class of the selected predictor (cis only, cis$+$trans, trans only); the dashed line is the null expectation $E|Z|=\sqrt{2/\pi}$, the dotted line marks the FDR $5\%$ significance boundary, and the horizontal bars are medians. (b) Number of significant genes (FDR$<0.05$) per tissue, stacked by predictor class, with the percentage that uses the trans channel shown above each bar. (c) Per-gene $|z_{\mathrm{TWAS}}|$ against the predicted-expression scale $\sigma_g$ (the denominator of the statistic): grey points are all tested genes, coloured points the significant genes by class. The total statistic carries no $\sigma_g$-driven inflation ($r(|z_{\mathrm{TWAS}}|,1/\sigma_g)\approx0$); after the trans $\sigma_g$ floor the trans channel retains only a weak residual dependence among significant hits ($r(|z_{\mathrm{trans}}|,1/\sigma_g)\approx0.15$). (d) Transcriptome-wide Manhattan of the best $p$ per gene across the seven tissues, with significant genes labelled by HGNC symbol.}
\label{fig:assoc}
\end{figure*}

To test whether the genetically predicted component of expression associates with disease, we combined each gene's predictor with the  CAD GWAS\citep{aragamDiscoverySystematicCharacterization2022} in a summary-statistic transcriptome-wide association (TWAS): a gene's standardised SNP weights are multiplied by the GWAS effect sizes and summed, normalised by the predicted-expression spread, to give a per-gene association statistic $z_{\mathrm{TWAS}}$ that decomposes additively into a cis and a trans component (Methods). Across the seven tissues this yields $2{,}407$ significant gene associations (FDR$<0.05$; $1{,}437$ unique genes; Fig.~\ref{fig:assoc}; Supplementary Table~S3). Of these, $1{,}572$ are modelled by a cis-only predictor, $714$ by a cis$+$trans predictor and $121$ by a trans-only predictor, so $835$ ($35\%$) use the trans information. The number of significant genes ranges from $236$ (Blood) to $510$ (AOR) per tissue, with the trans-using share between $24\%$ and $52\%$ (Fig.~\ref{fig:assoc}b). As expected, the associations recapitulate well-known CAD genes:\textit{ ADAMTS7, CDKN2B at 9p21, SORT1, CELSR2, PSRC1, PHACTR1, LPA, PCSK9, SLC22A3, PLG, TCF21 and EDNRA} all appear among the genome-wide peaks (Fig.~\ref{fig:assoc}d).

The trans-only associations depend entirely on the regulatory network and lack any cis-based anchor, which raises two specific concerns that we address directly. The first is that these genes might not reflect true signal, but instead form a weak residual class—predictors with only minimal correlation to expression that slip under the FDR threshold by chance under multiple testing, while otherwise behaving like nulls. Under the null, $z_{\mathrm{TWAS}}$ is standard normal, so the expected value of $|z_{\mathrm{TWAS}}|$ is $\sqrt{2/\pi}\approx0.80$. The second concern goes in the opposite direction: that a large statistic might arise purely from the denominator, since $z_{\mathrm{TWAS}}$ scales the genetic numerator by the predicted-expression standard deviation $\sigma_g$. An almost-degenerate predictor with very small $\sigma_g$ could thus produce a large $|z|$ even when the true signal is negligible. This issue did materialize for the trans channel: among trans-using genes, the trans component $|z_{\mathrm{trans}}|$ was correlated with $1/\sigma_g$ (Pearson $r\approx0.27$), indicating that predictors with small $\sigma_g$ inflated the trans signal. To address this, we imposed a minimum $\sigma_g$ threshold for the trans channel, deeming a trans-using predictor non-significant when its $\sigma_g$ falls below this floor, thereby removing the small-$\sigma_g$ trans hits most vulnerable to inflation. After applying this conservative filter, $121$ trans-only associations remain ($108$ distinct genes). The floor reduces this dependence but does not eliminate it entirely: in the remaining trans-using genes, the trans component shows only a weak correlation with $1/\sigma_g$ ($r\approx0.15$; Fig.~\ref{fig:assoc}c), much lower than the $r\approx0.27$ observed across all trans predictors prior to flooring, and the overall $|z_{\mathrm{TWAS}}|$ is essentially uncorrelated with $1/\sigma_g$ ($r\approx0$). Thus, the surviving signals are driven by predictors with moderate-to-large $\sigma_g$. The trans-only class is therefore small but robust: a conservatively curated set of network-only associations. These remaining genes exhibit nearly the same $|z_{\mathrm{TWAS}}|$ distribution (Fig.~\ref{fig:assoc}a), with trans-only genes closely matching cis-only genes in mean $|z|$ ($1.15$ vs $1.11$), and with a tail that extends well beyond the FDR threshold.

\subsection{Validation of the associations and their regulatory edges}\label{sec:negctrl}
The analysis above indicates that the associations, including the network-only ones, represent genuine statistical signal in the discovery data. To move from statistical signal to biological inference, we next consider several potential limitations. First, the regulatory edges that carry the trans signal are inferred from expression rather than observed, so they may not correspond to real molecular interactions. Second, even real edges, laid over a large and densely connected network, could
manufacture associations from topology alone, independently of the trait. Third, the associations are fit to a single discovery GWAS and may not generalise, and the trans channel in particular could be a mechanical echo of known loci. We address these with curated-interaction corroboration, a permutation negative control, and independent-cohort replication.
\begin{figure*}[htbp]
\centering
\includegraphics[width=\textwidth]{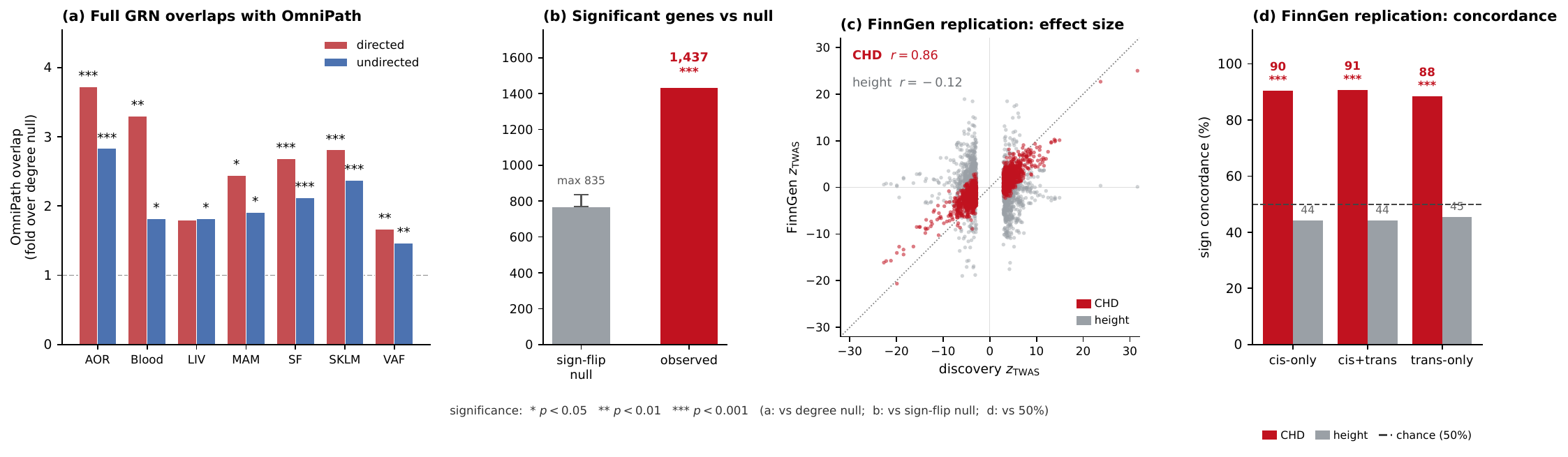}
\caption{Validation of the disease associations. \textbf{(a)} In each of the seven tissues, the full per-tissue GRN shows greater overlap with OmniPath interactions than a degree-preserving null model (which reassigns each edge’s endpoints to genes with matched OmniPath degree, preventing hubs from artificially increasing the overlap), for both exactly matching directed edges and undirected (either-orientation) matches. Directed overlap is significantly enriched in six of the seven tissues (AOR, Blood, MAM, SF, SKLM, and VAF; up to $3.8\times$ above the null). The dashed line indicates no enrichment, and asterisks indicate significant deviation from the null. \textbf{(b)} GWAS sign-flip negative control: the observed count of unique significant genes (red) compared with the permutation null (grey: bar for the mean, whisker to the maximum across 1000 sign-flip permutations); the observed value exceeds all permutations ($p<0.001$). \textbf{(c,d)} Associations reproduce in an independent cohort: we re-ran the identical TWAS in FinnGen R11 across all tested genes, changing only the summary statistics. \textbf{(c)} Discovery versus FinnGen $z_{\mathrm{TWAS}}$ among discovery-significant genes, for coronary heart disease (CHD, red) with FinnGen height as a negative control (grey); the dotted line is the identity, and the annotations report Pearson correlations (CHD $r=0.86$, height $r=-0.12$). \textbf{(d)} Concordance of effect direction in FinnGen with the discovery by predictor class, for CHD versus height; the dashed line marks chance level ($50\%$). Asterisks throughout indicate $*\,p<0.05$, $**\,p<0.01$, $***\,p<0.001$ (panel a versus the degree-preserving null, b versus the sign-flip null, d versus $50\%$).}
\label{fig:validation}
\end{figure*}

\paragraph{Overlap with curated molecular interactions.} To check that the inferred GRNs overlap with real regulatory biology rather than reconstruction noise, we compared their edges against OmniPath, a database of curated, directed molecular interactions \citep{tureiOmniPathGuidelinesGateway2016}, restricting to edges whose endpoints both appear in OmniPath. In every tissue the full GRN overlaps with OmniPath directed interactions more often than a degree-preserving null \citep{maslovSpecificityStabilityTopology2002} that fixes each gene's OmniPath connectivity, so hubs cannot inflate the overlap; the enrichment is
significant in six of the seven tissues (AOR, Blood, MAM, SF, SKLM and VAF; up to $3.8\times$ over null; Fig.~\ref{fig:validation}a, Supplementary Tables~S4--S6). So the inferred edges overlap curated interactions more than expected from random gene connectivity alone.


\paragraph{A GWAS sign-flip negative control.} To verify that our associations arise from the \emph{disease} GWAS signal itself, rather than from artifacts driven by a large, densely connected network, we designed a negative control targeting the only disease-specific input. Specifically, we permuted the GWAS $z$-scores by drawing an independent random sign ($\epsilon_s\in\{-1,+1\}$) for each SNP $s$ and applying that single sign vector across all genes and tissues, then reran the entire analysis pipeline; we repeated this 1,000 times, keeping the network structure, expression weights, and genotypes fixed (Methods). Under this procedure, all measured quantities converge to their null expectations (Fig.~\ref{fig:validation}b; Supplementary Table~S7): the $1{,}437$ distinct significant genes are reduced to a null mean of $768$ (a $1.9\times$ decrease), and the number of OmniPath-supported directed disease edges declines from $9$ to $2.4$ (a $3.8\times$ decrease). In every case, the original value exceeds the maximum observed across all $1{,}000$ sign-flips ($p<0.001$). Therefore, the size and connectivity of the network alone generate a notable baseline signal (the null mean, roughly half of the observed gene count), in part because some predictors are unaffected by sign flips; nevertheless, the true signal is about $1.9$ times this baseline and is larger than in all permutations, showing that a substantial portion depends on the concordance between regulators’ genetic effects and disease risk, rather than on network topology alone.

\paragraph{Replication in an independent cohort.} The null model evaluates artefacts arising from the network structure in the discovery dataset; an independent cohort instead tests artefacts specific to the discovery sample itself. We therefore repeated the same TWAS in FinnGen R11 coronary heart disease \citep{kurkiFinnGenProvidesGenetic2023}, a CAD cohort independent of the discovery set, evaluating all previously tested genes while only replacing the summary statistics. Across all tested gene–tissue pairs, the discovery and FinnGen associations are consistent genome-wide (Pearson $r=0.52$, $62\%$ sign concordance), and they align even more closely within the subset significant genes (Pearson $r=0.86$,, $90\%$ sign concordance; cis-only $90\%$, cis+trans $91\%$, trans-only $88\%$; all $p<10^{-18}$; Fig.~\ref{fig:validation}d, Supplementary Table~S8). One quarter of the discovery-significant associations ($592/2{,}399$) also reach significance under FinnGen’s own FDR threshold. This agreement is both directional and trait-specific: for height, used as a negative control, the same metrics drop to near random (genome-wide $r=-0.09$, $47\%$ sign concordance; discovery-significant $44\%$), despite height—being highly polygenic—having many significant TWAS genes in its own right. Thus, the observed replication primarily reflects shared CAD genetic architecture, rather than artifacts of the discovery sample or generic expression heritability.

\subsection{The disease-regulatory flux map}
\begin{figure*}[htbp]
\centering
\includegraphics[width=\textwidth]{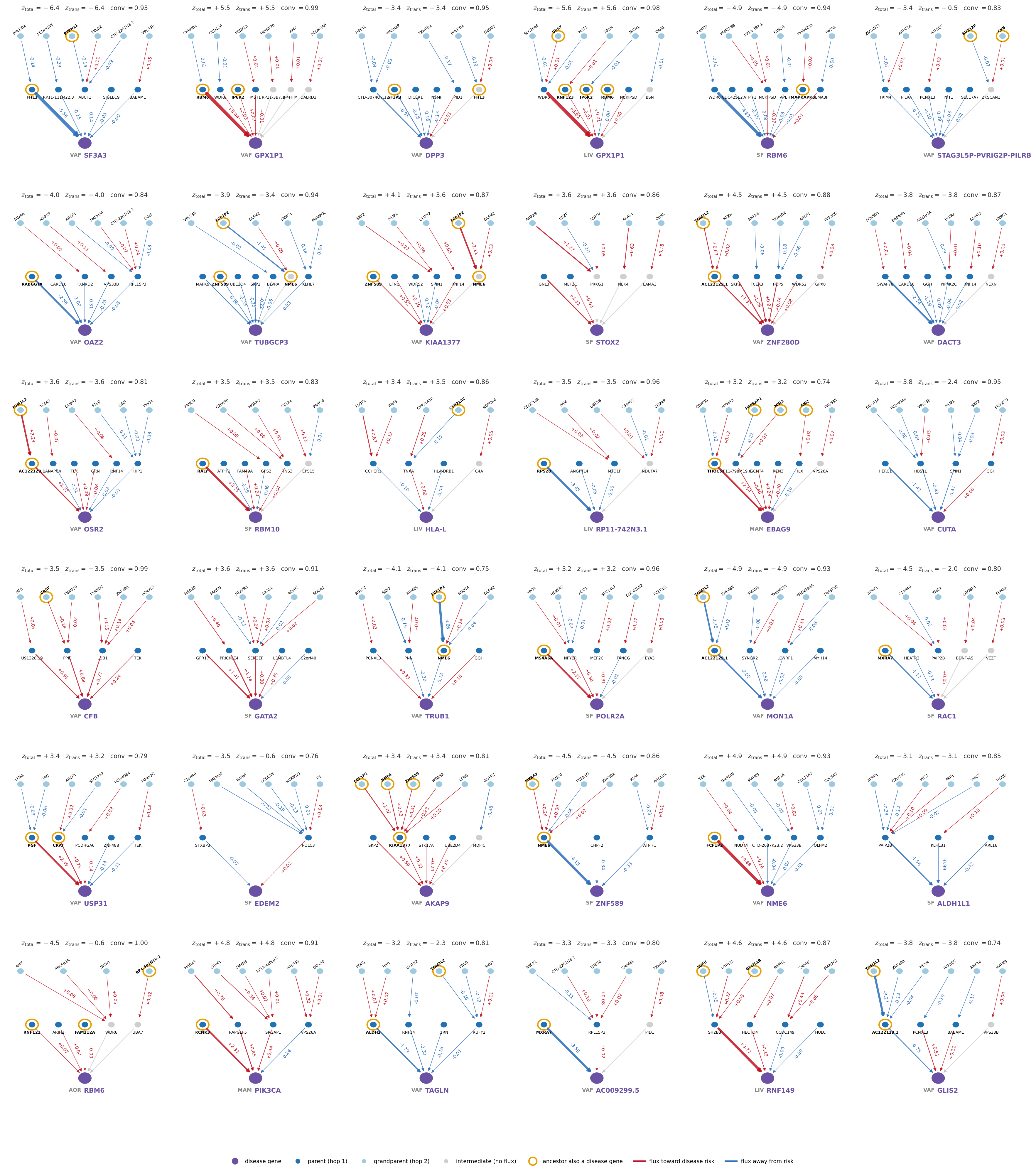}
\caption{The disease-regulatory flux map, shown as the two-hop regulatory cascade for the thirty-six most trans-driven CAD gene--tissue cases pooled across the seven tissues (the tissue is shown before each disease gene name), restricted to genes whose incoming regulator flux is convergent (net trans signal exceeding $0.7$ of the total incoming $\lvert\mathrm{flux}\rvert$). In each panel the disease gene (purple) is at the bottom and its regulators are drawn as the actual GRN cascade that
delivers the trans signal: grandparents (hop~2, light blue) regulate parents (hop~1, blue), which regulate the disease gene. A gold ring marks an ancestor that is itself a disease-significant gene; a grey node is an intermediate parent that routes a grandparent's signal but carries no flux of its own. Each edge carries the signed flux that its ancestor delivers to the disease gene: width is proportional to $\lvert\mathrm{flux}\rvert$, colour gives the sign (red, the genetics push the gene toward disease risk; blue, away), and the signed value is labelled on the edge. Per gene these contributions sum exactly to $\ztrans$; the gene's total and trans association
($z_{\mathrm{total}}$, $z_{\mathrm{trans}}$) and its incoming convergence (conv $=\lvert\ztrans\rvert/$in-flux) are shown above each panel (Eq.~\eqref{eq:flux}). Grandparents typically converge onto a few parents, and hub regulators recur across genes.}
\label{fig:flux}
\end{figure*}

\begin{figure*}[htbp]
\centering
\includegraphics[width=\textwidth]{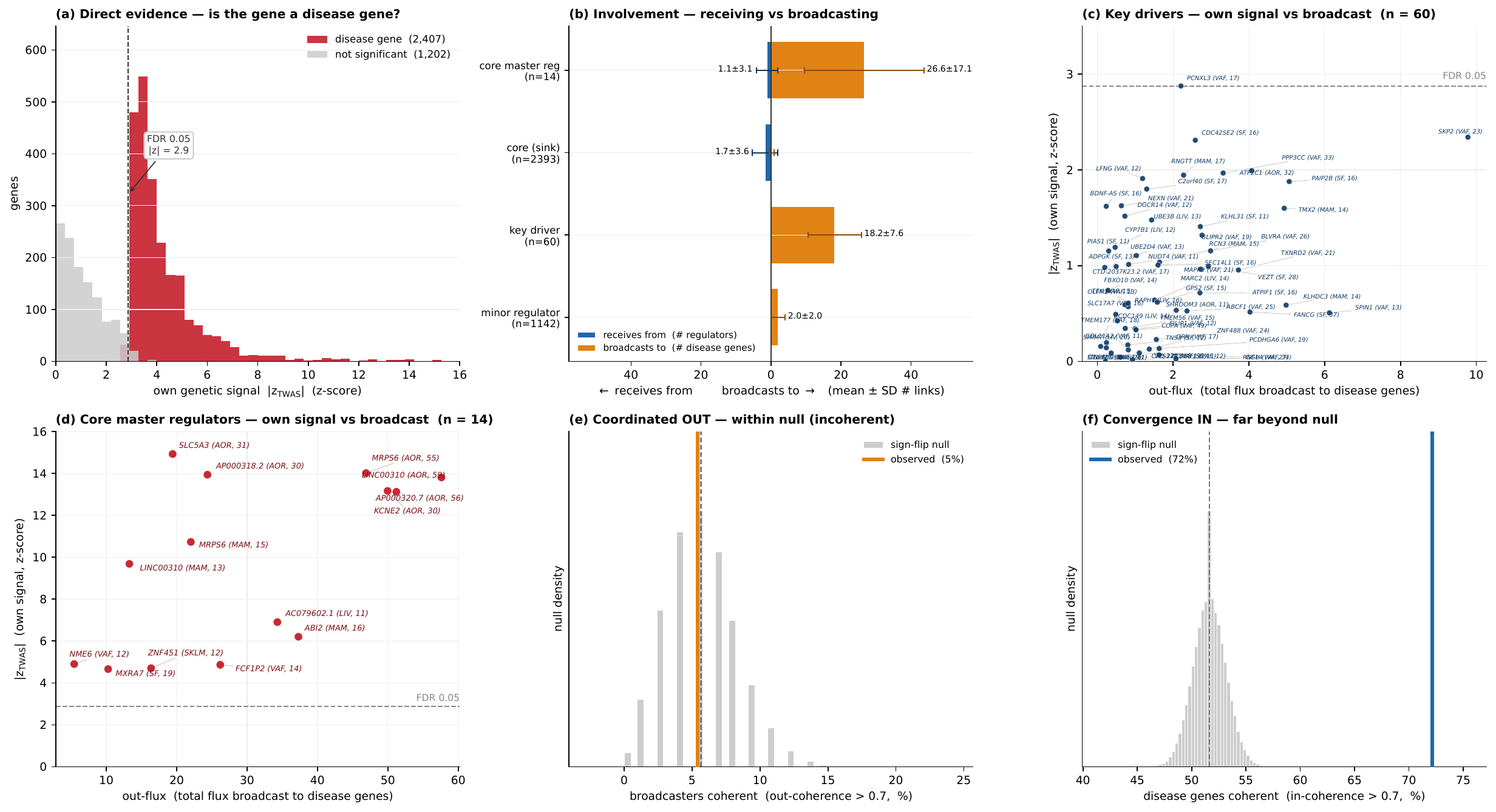}
\caption{Complementary views of the disease flux map. Genes are aggregated across the seven tissues as gene–tissue instances: a gene that is significant (or functions as a regulator) in multiple tissues contributes one instance per tissue ($3{,}609$ instances total—$2{,}407$ disease-gene instances from $1{,}437$ distinct genes, plus $1{,}202$ instances of non-significant regulators). \textbf{(a)}~\emph{Direct evidence}: intrinsic genetic signal $\lvert z_{\mathrm{TWAS}}\rvert$ (z-score) for the $2{,}407$ disease-gene instances (red) versus the $1{,}202$ regulators that exert trans effects but are not themselves significant (grey). The dashed line marks the FDR-$0.05$ threshold (a small number of genes with $\lvert z\rvert$ beyond the axis range are omitted). \textbf{(b)}~\emph{Involvement}: incoming regulation (number of regulators, blue) versus outgoing influence (number of disease-gene targets, orange) across the four gene categories; bars represent class means $\pm$ SD over instances. \textbf{(c)}~Key drivers (non-significant broadcasters) and \textbf{(d)}~core master regulators (significant broadcasters) are each shown with their own genetic signal $\lvert z_{\mathrm{TWAS}}\rvert$ (z-score, vertical axis) against out-flux (horizontal axis), the total flux a gene broadcasts to disease genes. Key drivers occur exclusively below the FDR threshold and master regulators exclusively above it—non-overlapping $\lvert z\rvert$ ranges—so the vertical axis scale in each panel is specific to that group. All key drivers are labelled in (c) and all master regulators in (d). \textbf{(e)}~\emph{Coordination} of broadcasting direction: the proportion of broadcasters with highly coherent outputs ($>0.7$) falls within the sign-flip null (coloured line, observed; grey, null distribution). \textbf{(f)}~\emph{Convergence} of receiving direction: the ($z=14\sigma$, $200{,}000$ permutations). Together, panels (e,f) indicate that signal converges onto disease genes (many$\to$one) without exhibiting strong, coherent divergence from individual regulators (one$\to$many).}
\label{fig:readouts}
\end{figure*}

Our associations, their concordance with curated interaction databases, and their replication in an independent cohort collectively indicate that the trans channel captures genuine disease-related signals. We therefore leverage this channel to construct an alternative disease-gene prioritization framework that explicitly traces how genetic variation exerts its effects on disease through the regulatory network, which we call the disease-regulatory flux. The CAD flux map spans $2{,}407$ disease-significant genes across the seven tissues (Fig.~\ref{fig:assoc}) and the $1{,}202$ upstream regulators that feed trans signal into them without reaching genome-wide significance themselves. Genes whose association is wholly or partly trans (the trans-only and cis–trans classes) draw that signal from these regulators.

In this way, the flux map expands the set of disease-relevant genes to include regulatory genes that drive the observed signals, providing a novel strategy in the TWAS context. By construction, the trans component of a gene’s association is exactly the sum of the signed contributions from upstream regulators whose cis-eQTLs propagate through the network to that gene, $\sum_A \mathrm{flux}(A\to g)=\ztrans(g)$ (Supplementary Table~S9). This representation allows us to backpropagate contributions from disease-significant genes to their contributing regulators. Figure~\ref{fig:flux} illustrates examples of disease-regulatory flux for genes with the largest trans $z$-scores across tissues and interactive largest-connected-component flux maps for all seven tissues are provided as Interactive Supplementary Figures~1--7.. The resulting map can be used to explore and interpret the architecture and mechanisms underlying complex traits. Using this framework, we examined several distinct axes for CAD (Fig.~\ref{fig:readouts}).

\paragraph{Involvement.} We use the flux map to characterize how each gene that shapes disease relevance both receives and routes risk toward disease. Every gene simultaneously plays a \emph{receiving} role (being influenced by local and distal genetic variants) and a \emph{broadcasting} role (propagating local and distal genetic effects to other genes), and positioning genes along these two dimensions separates them into four categories (Fig.~\ref{fig:readouts}b). The majority of CAD genes function as \emph{sinks} ($2{,}393$ in total): they receive input from only a small number of regulators (mean $1.7$) and emit almost no outgoing signal. A small group of \emph{core master regulators} ($14$ genes) is both disease-significant \emph{and} highly broadcasting (on average they regulate $27$ disease targets, with input from only $1$ regulator on average; Fig.~\ref{fig:readouts}d). Among regulators that are not themselves disease-significant, $60$ \emph{key drivers} send signals to many disease genes (mean $18$; Fig.~\ref{fig:readouts}c), whereas the remaining $1{,}142$ \emph{minor regulators} each connect to only a few disease genes (per-gene annotations in Supplementary Table~S10). Thus, broadcasting extent is largely decoupled from a gene’s own disease significance: numerous genes transmit disease-related signals through the network without individually surpassing the disease-gene threshold (this connectivity is made explicit in the cascade view in Fig.~\ref{fig:flux}.

\paragraph{Broadcasting is not coherent.} We next examined whether the flux is directionally coordinated and found a strong asymmetry between the two directions. The outputs of a broadcaster are \emph{not} coherent: only $5\%$ of broadcasters predominantly influence their disease targets in a single direction, essentially matching the $6\%$ expected under a sign-flip null (Fig.~\ref{fig:readouts}e).

\paragraph{Regulation converges coherently.} In the incoming direction, the pattern is reversed. Fully $72\%$ of disease genes show coherent regulation, with most of their incoming effects pointing in the same direction, far exceeding the $52\%$ expected under the corresponding null (Fig.~\ref{fig:readouts}f; $z=14\sigma$, $p<5\times10^{-6}$ across $200{,}000$ permutations). We can only hypothesize about why the two directions behave differently. One possible explanation is that coherence is itself a prerequisite for a gene to exert a detectable disease effect: a regulator sends signals to many targets that it may influence for unrelated reasons, so its outgoing effects need not be aligned with respect to disease, whereas a gene becomes recognizably disease-associated specifically when many of its incoming influences happen to line up in the same direction, and the disease signal appears as a byproduct of that convergence. Under this interpretation, disease-implicated genes are the loci at which regulation converges to the targets rather than the upstream broadcasters, and the \emph{distinct, independent} cis-eQTLs carried by many regulators are what permit aligned inputs to sum instead of cancel. Regardless of mechanism, the asymmetry we detect—outgoing flux being incoherent while incoming flux converges—resembles the omnigenic view, where numerous individually weak, not necessarily aligned effects accumulate on a core gene. 

\subsection{Testing the prioritised genes for cardiovascular relevance}\label{sec:funcmethods}
\begin{figure*}[htbp]
\centering
\includegraphics[width=\textwidth]{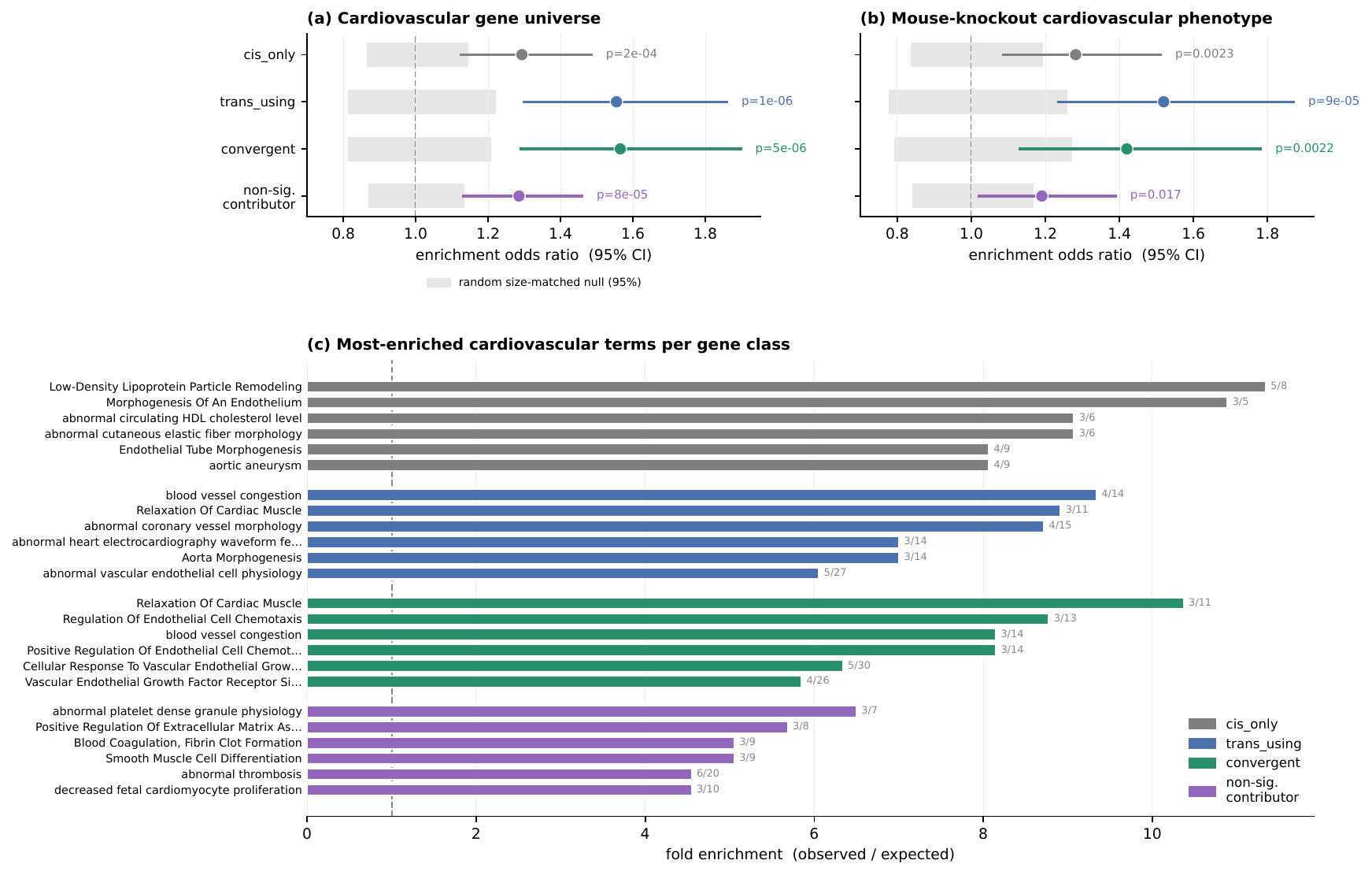}
\caption{Enrichment odds ratios (points) with $95\%$ confidence intervals (lines) for four gene classes—cis-only, trans-using (associations leveraging the trans channel), convergent (coherent input from multiple regulators) and non-significant contributors (regulators highlighted by the network but not individually reaching significance)—relative to the tested-gene background, for \textbf{(a)}~a curated cardiovascular gene universe (mouse-knockout phenotypes, Gene Ontology, Reactome, KEGG and curated disease annotations) and \textbf{(b)}~mouse-knockout cardiovascular phenotypes alone, providing a readout independent of human genetic data. The dashed horizontal line indicates no enrichment (OR~$=1$); the grey band behind each class represents the $95\%$ range for random, size-matched gene sets (centred at OR~$=1$); and $p$-values are from one-sided Fisher exact tests. All four classes show enrichment on both axes and exceed the random null. \textbf{(c)}~Most-enriched individual cardiovascular terms per class: bar length denotes fold enrichment (observed$/$expected), bar colour denotes the class, and each bar is annotated with the term and, in the form $k/K$, the number of class genes in the term ($k$) out of the term’s tested size ($K$)—for example, $5/8$ means that five of the eight tested genes for that term belong to the class. Cis-only genes are chiefly enriched for lipid-related processes, whereas trans-using and convergent classes are dominated by vascular and cardiac processes, and the non-significant contributors are primarily linked to coagulation, thrombosis and smooth-muscle processes.}
\label{fig:functional}
\end{figure*}

The map prioritises disease genes based on genetic evidence; we then asked whether the highlighted genes and regulators are cardiovascular according to criteria that do not rely on the CAD GWAS itself. We evaluated four sets: disease genes that are genome-wide significant via the cis channel only (\emph{cis-only}); disease genes whose association involves the trans channel (\emph{trans-using}, i.e.\ cis$+$trans or trans-only); \emph{convergent} disease genes, which consistently integrate signals from multiple upstream regulators; and the \emph{non-significant contributors}—regulators that deliver trans signal to disease genes but do not themselves reach significance. For each set, we tested enrichment relative to all tested genes against a curated cardiovascular gene universe ($5{,}110$ of $16{,}769$ tested genes, $30\%$) and, as an orthogonal readout independent of human genetics, mouse-knockout genes with cardiovascular phenotypes ($2{,}872$ genes, $17\%$), using a one-sided Fisher’s exact test (Fig.~\ref{fig:functional}). Details of universe construction, group definitions and individual term-level tests are provided in Supplementary Section~S6. The chosen cardiovascular gene sets, the resulting universe, per-group enrichment statistics and top terms are summarised in Supplementary Tables~S11--S14, and the Enrichr libraries used are archived alongside these tables.

\paragraph{Disease genes and their regulators are cardiovascular, including in mouse knockouts.} Every gene category highlighted by the map is enriched for cardiovascular roles relative to the tested-gene background: cis-only disease genes (odds ratio $1.29$, $p=2\times10^{-4}$), trans-using genes—whose associations depend on the trans channel (OR~$1.55$, $p=1\times10^{-6}$)—convergent genes, which integrate signals from multiple upstream regulators (OR~$1.57$, $p=5\times10^{-6}$), and even the non-significant “contributors,” regulators that the network implicates despite not individually reaching significance (OR~$1.29$, $p=8\times10^{-5}$; Fig.~\ref{fig:functional}a). In contrast, random gene sets matched for size yield odds ratios centred around $1$ (grey band; Fig.~\ref{fig:functional}a,b), showing that all four categories are well beyond what would be expected by chance. This enrichment also appears when considering cardiovascular phenotypes in mouse knockouts, an orthogonal assay independent of human genetics (cis-only OR~$1.28$; trans-using OR~$1.52$; convergent OR~$1.42$; contributors OR~$1.19$; all $p<0.02$; Fig.~\ref{fig:functional}b), indicating that it is not simply a consequence of the original GWAS design. The trans-using and convergent categories show the strongest enrichment, comparable to or exceeding the canonical cis-anchored genes: the map therefore propagates cardiovascular relevance across the network, capturing genes and regulators that a cis-only analysis would overlook.
\paragraph{Distinct gene classes reflect complementary aspects of cardiovascular biology.} The most strongly enriched individual annotations vary by gene class (Fig.~\ref{fig:functional}c; Supplementary Table~S14): cis-only genes are chiefly associated with lipid and lipoprotein traits (low-density-lipoprotein particle remodelling, circulating-cholesterol level), the trans-using and convergent classes are enriched for vessel, myocardial and endothelial functions (coronary-vessel morphology, cardiac-muscle relaxation, endothelial chemotaxis), and the non-significant contributors are linked to thrombotic and vessel-wall biology (platelet function, fibrin-clot formation, smooth-muscle differentiation, extracellular-matrix assembly). While the cis-anchored genes capture the lipid axis of coronary disease that typically dominates cis-TWAS, the network highlights complementary vascular, endothelial and thrombotic axes. These class-specific distinctions are descriptive—each term involves only a small number of genes and none remains individually significant after multiple-testing correction—but the overall pattern is consistent across classes.

\paragraph{The signal is distributed, not driven by a single pathway.} The enrichment arises from the combined set of cardiovascular annotations: no single ontology term remains significant after multiple-testing correction, so the signal cannot be attributed to one dominant process. In line with this, the regulators converging on a given disease gene do not form a tight functional module—their pairwise pathway co-membership is comparable to that of size-matched random gene sets—implying that convergence reflects the integration of functionally heterogeneous inputs rather than the reinforcement of a single pathway.

\section{Discussion}

\begin{table*}[htbp]
\centering
\footnotesize
\setlength{\tabcolsep}{8pt}\renewcommand{\arraystretch}{1.4}
\arrayrulecolor{tabhead}
\begin{tabular}{@{}>{\raggedright\arraybackslash}p{3.05cm}
                   >{\raggedright\arraybackslash}p{5.25cm}
                   >{\raggedright\arraybackslash}p{5.25cm}@{}}
\arrayrulecolor{tabhead}\specialrule{1pt}{0pt}{0pt}
\rowcolor{tabhead}
{\color{white}\textbf{Method family}} &
{\color{white}\textbf{What it shares with our approach}} &
{\color{white}\textbf{Key differences}} \\
\specialrule{1pt}{0pt}{0pt}

\textbf{Variance / heritability partitioning}\newline{\scriptsize S-LDSC
\citep{reprogenconsortiumPartitioningHeritabilityFunctional2015}; omnigenic trans-$h^2$ \citep{liuTransEffectsGene2019}} & Quantify, in aggregate, how much trait or expression variance is carried by regulatory variation: S-LDSC partitions trait heritability across functional annotations, and the omnigenic estimator gives the trans share of expression heritability. & Return genome-wide variance shares: a per-annotation vector (S-LDSC) or a single cis/trans fraction (omnigenic $h^2$). Neither has per-gene resolution, a cis/trans split for an individual gene, or named relaying regulators. \\[2pt]

\rowcolor{tabalt}
\textbf{Trans-aware association}\newline{\scriptsize MOSTWAS \citep{bhattacharyaMOSTWASMultiOmicStrategies2021}; GBAT \citep{liuGBATGenebasedAssociation2020}; trans-PCO \citep{wangTranseQTLMappingGene2024}; BGW-TWAS \citep{luninghamBayesianGenomewideTWAS2020}} & Also bring distal (trans) genetic signal to bear on gene-level regulatory effects to recover power: through imputed mediators, cis-predicted regulator genes, module principal components, or genome-wide eQTL selection. & None reconstructs a causal gene-regulatory network or carries a regulator's cis-eQTLs along it \emph{inside} the expression predictor, so none returns a signed per-regulator flux or an additive own/network split of each association. What each does return differs: MOSTWAS names local mediators (TFs, miRNAs, CpGs) via mediation analysis; GBAT names candidate upstream genes through cis-predicted gene-to-gene tests; BGW-TWAS jointly selects cis and trans SNPs under separate priors and can attribute an association to trans; trans-PCO maps SNP$\to$co-expression-module trans-eQTLs without naming the driver genes. \\[2pt]

\textbf{Gene-network causal inference}\newline{\scriptsize BN-GWAS
\citep{yinEstimationCausalEffects2024}} & Also learns a directed gene--gene--phenotype causal network and splits each gene's effect on the trait into a direct and a mediated part. & Learns the network as a separate constraint-based structure-learning step (PC-simple for gene--phenotype, PC for gene--gene edges) run on
already-imputed expression, recovering a partially oriented CPDAG rather than a fully oriented DAG; it does not fold a regulator's cis-eQTLs into the target gene's predictor, and its direct/mediated split is of the causal \emph{effect}, not an additive cis/trans decomposition of the \emph{association} (distal signal
enters only as separately supplied trans-eQTL data). \\

\specialrule{1pt}{0pt}{0pt}
\rowcolor{tabours}
\textbf{Disease-regulatory flux map}\newline{\scriptsize (this work)} &
\multicolumn{2}{>{\columncolor{tabours}}p{10.95cm}@{}}{Reconstructs a cis-anchored causal GRN directly from reference genotype and expression (each
edge oriented by the source gene's own cis-eQTL) and propagates each regulator's cis-eQTLs along that DAG \emph{inside} the expression predictor.
Every association then splits additively into an own (cis) and a network (trans) component, and each regulator$\to$disease-gene link carries a \emph{signed
flux}: a per-gene, per-regulator account of where the signal comes from and in which direction it pushes risk.} \\
\specialrule{1pt}{0pt}{0pt}
\end{tabular}
\arrayrulecolor{black}
\caption{The disease-regulatory flux map alongside trans- and network-aware approaches. For each family we name representative methods, what they
\emph{share} with the flux map, and how they differ. Existing methods return an aggregate variance share, a distal-signal-augmented gene test, or a causal
network learned downstream of imputation; the flux map instead propagates each regulator's cis-eQTLs along an inferred causal GRN inside the predictor,
yielding a per-gene cis/trans split and named, signed regulator-to-gene links.}
\label{tab:comparing-methods}
\end{table*}

The disease-regulatory flux map resolves a gene's summary-statistic association into where its signal comes from: it gives each gene an own (cis) and a network (trans) coordinate and assigns every regulator-to-disease-gene link a signed weight. The result is a directed, weighted account of how a trait's genetic signal is distributed across an inferred regulatory network, the flow object that the omnigenic model named but did not construct \citep{liuTransEffectsGene2019}. This is a structured, prioritised output rather than a flat hit list: genes can be ranked not only by their own association but also by the signal they receive from or relay to others; it names the specific regulators behind each disease gene; and it surfaces regulators whose effects are individually weak and hard to detect,
and that need not be functionally related to one another, yet whose coordinated, same-direction action converges on shared disease genes, carrying signal that none shows on its own. Prioritisation can then be guided by the regulatory wiring (which regulator influences which disease gene, and in what direction) rather than by gene-level associations alone. This provides a concrete entry point for downstream experimental or causal-inference efforts, such as drug target prioritisation and repurposing or CRISPR screening.

This work sits alongside a growing set of methods that look beyond the cis window using trans or network information (Table~\ref{tab:comparing-methods}); what mainly
separates it from them is the resolution at which the trans signal is placed. Variance-partitioning approaches (S-LDSC \citep{reprogenconsortiumPartitioningHeritabilityFunctional2015}; the omnigenic trans-$h^2$ estimator \citep{liuTransEffectsGene2019}) measure, genome-wide, how
much heritability is regulatory (S-LDSC, across functional annotations) or specifically trans-acting (the omnigenic estimator, as a cis/trans split of
expression heritability), but only in aggregate, not resolved to individual genes or to the regulators that relay the signal. Trans-aware association methods
(MOSTWAS \citep{bhattacharyaMOSTWASMultiOmicStrategies2021}; GBAT \citep{liuGBATGenebasedAssociation2020}; trans-PCO \citep{wangTranseQTLMappingGene2024}; BGW-TWAS \citep{luninghamBayesianGenomewideTWAS2020}) bring distal signal into the gene-level test, but none reconstructs a causal regulatory network linking regulators to their targets, so the recovered
trans signal is not returned as a directed, signed set of per-regulator contributions to a gene's association. The closest approach, BN-GWAS
\citep{yinEstimationCausalEffects2024}, does learn a directed gene--gene--phenotype causal network and separates each gene's direct from its mediated effect on the trait, but it builds that network by constraint-based structure learning on already-imputed expression (downstream of, rather than inside, the expression predictor), so a regulator's cis-eQTLs are never carried into the prediction of its targets. We take the complementary route: we reconstruct a cis-anchored causal network directly from reference genotype and expression, and propagate each regulator's cis-eQTLs along it \emph{inside} the predictor, so that every gene's disease association separates into an own (cis) and a network (trans) part and each regulator-to-target link carries a signed contribution, placing the trans signal at the level of individual genes and named, directed regulator-to-gene links.

Applied to CAD in seven STARNET tissues, the map recovers curated molecular interactions above a degree-preserving null, replicates in an independent cohort, and highlights a category of disease genes whose association signal is assembled in trans. Considering each gene along four axes, we observe an asymmetry: a gene’s outgoing flux to its targets is, on average, no more directionally organized than expected under a sign-flip null, whereas the incoming flux onto disease genes is highly convergent ($72\%$ vs.\ $52\%$). All of the classes we tests for enrichment (the map—cis-only, trans-using, convergent and non significant contributors) are enriched for cardiovascular annotations, including a mouse-knockout phenotype readout that is independent of human genetic data, with the trans-using and convergent classes performing at least as well as the traditionally cis-anchored genes. We offer these as observations about CAD’s regulatory architecture, in line with the omnigenic view that many weak, individually incoherent inputs aggregate onto core genes, while acknowledging that they do not constitute formal proof of that model.

The trans channel makes only a modest contribution to predicting expression: predictors that rely exclusively on trans achieve low leave-one-out $R^2$ (mean $<0.05$; Fig.~\ref{fig:pred}), far below those based on cis. This does not diminish the validity of the trans associations, because predicting expression and testing trait association are fundamentally different objectives. The trans statistic $\ztrans$ is defined as a weighted sum of a gene’s upstream regulators’ cis-eQTL effects, evaluated against the GWAS (Eq.~\ref{eq:twasz}); it tests whether the genetically grounded, network-propagated portion of a gene’s regulation tracks with disease risk, rather than whether it recovers the gene’s overall expression level. Since most of a transcript’s variance arises from cis and non-genetic influences, the trans-genetic component is relatively small—small enough to lead to low predictive $R^2$, yet, when aggregated over many regulatory variants and evaluated in a large, well-powered GWAS, still capable of generating a genuine association. This mirrors the well-known scenario in which many individual variants, each explaining a tiny fraction of trait variance, together form a strong polygenic signal.

That the trans associations reflect genuine disease biology rather than overfitting to the discovery cohort is supported by independent replication: the trans component shows >$90\%$ sign concordance between STARNET and FinnGen and is flat for height, used here as a negative control (Fig.~\ref{fig:validation}). This pattern is incompatible with simple overfitting to the STARNET expression data. Accordingly, we interpret the trans channel not as an expression-imputation
procedure but as a network-based strategy for prioritising genes downstream of disease-associated regulatory variants—a way to interrogate omnigenic architecture rather than to predict transcript levels.

A key unresolved issue is whether a gene identified via trans effects owes its disease relevance solely to its upstream regulators, rather than to its own functional role. Several observations argue against this. First, the upstream regulators themselves are usually not genome-wide significant for disease: they convey trans signal to their targets without individually exceeding genome-wide thresholds. As a result, a trans-identified gene rarely has a single
disease-significant “ancestor” that it merely proxies; instead, its signal arises from the aggregate contribution of many individually sub-significant regulators. Consistent with this, the identified genes are enriched for cardiovascular annotations, which would be unlikely if they were just arbitrary downstream targets of disease-implicated regulators; moreover, their trans effects replicate in direction in an independent cohort, which cannot be accounted for by shared functional annotation alone. We therefore attribute disease relevance to the trans-identified genes themselves, while recognising that, by construction, a gene’s trans signal cannot be cleanly separated from its regulatory context.

\paragraph{Limitations.} The regulatory network is inferred from a finite number of samples—on the order of $500$ individuals per tissue—and its edges represent statistically inferred causal relationships rather than experimentally validated ones. Consequently, our flux map is only as reliable as the underlying inferred network, which in turn depends heavily on data quality as well as sample size. We have argued that, in STARNET, the evidence indicates that the trans channel is modest in magnitude but genuinely present. Our current work is restricted to tissue-specific analyses, whereas real regulatory processes may also include cross-tissue interactions. The acyclicity assumption excludes the feedback loops that exist in actual regulatory systems, but these loops cannot be resolved from cross-sectional, single-time-point expression data in any case. The flux weights are associative: a large flux from a regulator to a disease gene should be interpreted as a hypothesis about a relayed signal, not as proof that the regulator is a causal driver of disease. The functional enrichments we observe are statistically significant but modest in size (odds ratios $\approx 1.3$--$1.6$), and they rely on curated annotations that are themselves incomplete and subject to study bias. Finally, all results pertain to a single trait in a single cohort of cardiometabolic tissues; whether a similar convergence pattern is observed for other traits and tissues remains an open question.

\section{Conclusion}
The disease-regulatory flux map transforms each gene’s TWAS association into both an intrinsic and a network-based coordinate, together with a set of signed regulator–to–gene edges, thereby operationalising the omnigenic flow object for a specific trait. In the case of CAD, it uncovers a consistent asymmetry: outward broadcasting is diffuse and non-coherent, whereas regulatory input converges in an orderly fashion onto disease genes, and these convergent targets are functionally cardiovascular according to criteria that are independent of the GWAS. By pinpointing where each gene’s disease signal originates and tracing its propagation through the regulatory network, the map supplements gene-level association with a directional, regulator-specific description.

\section*{Data availability}

\noindent\textbf{Reference cohort (STARNET; dbGaP).} Individual-level genotype data and multi-tissue
RNA-seq from STARNET are accessible via Authorized Access under accession \textbf{phs001203.v4.p1}
\href{https://www.ncbi.nlm.nih.gov/projects/gap/cgi-bin/study.cgi?study_id=phs001203.v4.p1}{(link)};
use requires an approved Data Use Certification. 

\smallskip
\noindent\textbf{CAD discovery GWAS.} Summary statistics for coronary artery disease are taken from
the CARDIoGRAMplusC4D meta-analysis \citep{aragamDiscoverySystematicCharacterization2022} and can be
downloaded from the GWAS Catalog under accession
\href{https://www.ebi.ac.uk/gwas/studies/GCST90132314}{GCST90132314}.

\smallskip
\noindent\textbf{Replication and negative-control GWAS.} Summary statistics for coronary heart
disease (\texttt{I9\_CHD}) and height (\texttt{HEIGHT\_IRN}) are from the FinnGen R11 public
release \citep{kurkiFinnGenProvidesGenetic2023}
(\href{https://www.finngen.fi/en/access_results}{finngen.fi/en/access\_results}; per-endpoint datasets
\texttt{finngen\_R11\_I9\_CHD} and \texttt{finngen\_R11\_HEIGHT\_IRN}).

\smallskip
\noindent\textbf{Curated interactions and annotations.} Directed molecular interaction data are
sourced from OmniPath \citep{tureiOmniPathGuidelinesGateway2016} (\href{https://omnipathdb.org}{omnipathdb.org}). Gene
annotations and symbols follow GENCODE v19
(\href{https://www.gencodegenes.org/human/release_19.html}{release 19}). Cardiovascular gene-set
collections were obtained from Enrichr \citep{chenEnrichrInteractiveCollaborative2013,kuleshovEnrichrComprehensiveGene2016}
(MGI Mammalian Phenotype Level~4 2021, GO Biological Process 2023, Reactome 2022, KEGG 2021 Human,
and Jensen DISEASES; \href{https://maayanlab.cloud/Enrichr/\#libraries}{maayanlab.cloud/Enrichr}).

\smallskip
\noindent\textbf{Derived results.} All summarized outputs underlying the study---including the inferred
tissue-specific networks, per-gene expression-model summaries, TWAS association statistics, flux
edges, and enrichment analyses---are provided as Supplementary Tables, along with interactive
tissue-level flux maps (Interactive Supplementary Figures~1--7), and are deposited at Zenodo
(DOI: \href{https://doi.org/10.5281/zenodo.21106390}{10.5281/zenodo.21106390}).

\section*{Code availability}

\noindent The disease-regulatory-flux workflow—including network reconstruction, channel-aware
expression modeling, summary-statistics-based TWAS, and flux decomposition (implemented in Nextflow and Python)—is available at
\href{https://github.com/guutama/disease-regulatory-flux}{github.com/guutama/disease-regulatory-flux};
the exact version used here is archived on Zenodo
(DOI: \href{https://doi.org/10.5281/zenodo.21221662}{10.5281/zenodo.21221662}).
Gene-regulatory networks were inferred using the Findr C library
\citep{wangEfficientAccurateCausal2017,wangHighDimensionalBayesianNetwork2019}
(\href{https://github.com/lingfeiwang/findr}{github.com/lingfeiwang/findr}), and Bayesian expression
models were fitted with NumPyro \citep{binghamPyroDeepUniversal2018}.
.

\section*{Competing interests}
The authors declare no competing interests.

\section*{Acknowledgments}
\noindent\textbf{Funding.}
This work was supported by the Research Council of Norway (project numbers 312045, 331725) and the European Union Horizon Europe (European Innovation Council) programme (grant agreement number 101115381), and the L. Meltzers Høyskolefond.  J.L.M.B. acknowledges support from the Swedish Research Council (2018-02529 and 2022-00734), the Swedish Heart Lung Foundation (2017-0265 and 2020-0207), the Leducq Foundation AtheroGen (22CVD04) and PlaqOmics(18CVD02) consortia, the National Institute of Health–National Heart Lung Blood Institute (NIH/NHLBI; R01HL164577, R01HL148167, R01HL148239, R01HL166428 and R01HL168174), the American Heart Association transformational Project Award 19TPA34910021 and the CMD AMP fNIH program. 

\noindent\textbf{Resources.}
Computations were performed on resources provided by Sigma2-the National Infrastructure for High Performance Computing and Data Storage in Norway  (project number NN1015K).


\end{document}